\documentclass[twocolumn,english,showpacs,aps,prl]{revtex4}
\usepackage[T1]{fontenc}
\usepackage[latin1]{inputenc}
\usepackage{amsmath}
\usepackage{graphicx}
\usepackage{amssymb}

\makeatletter
\usepackage{graphicx}
\usepackage{amsfonts}

\usepackage{babel}
\makeatother
\begin{document}

\title{Coherent behavior of balls in a vibrated box }

\author{Y. Garrabos$^{\left(1\right)}$, P. Evesque$^{\left(2\right)}$,
F. Palencia$^{\left(1\right)}$, C. Lecoutre$^{\left(1\right)}$,
and D. Beysens$^{\left(3\right)}$}

\affiliation{$^{\left(1\right)}$ESEME-CNRS, ICMCB-UPR 9048, Université Bordeaux
I, 87 avenue du Docteur Albert Schweitzer, F-33608 Pessac France;}

\affiliation{$^{\left(2\right)}$LMSSMat-CNRS-UMR 8579, Ecole Centrale Paris,
F-92295 Châteney-Malabry France;}

\affiliation{$^{\left(3\right)}$ESEME-CEA, SBT, CEA-Grenoble, F-38054 Grenoble
Cedex 9, France.}

\date{revised 28 July, 2004}

\begin{abstract}
We report observations on very low density limit of one and two balls,
vibrated in a box, showing a coherent behavior along a direction parallel
to the vibration. This ball behavior causes a significant reduction
of the phase space dimension of this billiard-like system. We believe
this is because the lowest dissipation process along a non-ergodic
orbit eliminates ball rotation and freezes transverse velocity fluctuations.
>From a two-ball experiment performed under low-gravity conditions,
we introduce a {}``laser-like'' ball system as a prototype of a
new dynamical model for very low density granular matter at nonequilibrium
steady state.
\end{abstract}


\maketitle
The present letter starts with the experimental study of the dynamical
behavior of a single ball vibrated in a three-dimensional (3D) box.
It can be viewed as a 3D experimental version of accelerator models
of particle physics impacting oscillating heavy objects, and vibrating
billiard type systems, where the physics of ergodicity, i.e., filling
of the available phase space by stochastic motion, can be examined
from the bouncing ball models \cite{Lichtenberg 1980,Holmes 1982,Lichtenberg 1992}.
Figs. 1a and 1b give schematic presentation of the two most known
accelerator models of one bouncing ball, the so-called Pustylnikov
and Ulam \underbar{}versions of the Fermi acceleration mechanism (see
\cite{Lichtenberg 1980} and references therein for details). In the
Pustylnikov version, the ball moves freely above the vibrating wall,
under a constant acceleration (here the $g_{0}$ Earth's gravity acceleration),
while in the Ulam version, the particle moves with constant velocity
between impacts with two walls - one vibrating and one fixed.

It can be studied also as the ultimate limit for a forced dilute granular
gas \cite{Olafsen 1999}, when grain-grain collisions are negligible.
This can occur, for example, in a cubic cell when the small amount
$N$ of grains only covers a very small fraction of one vibrating
wall surface. Therefore, the grain mean free path $l_{g}$ between
two grain-grain collisions is much larger than the cell size $L$,
corresponding to the so-called Knudsen-like regime \cite{Evesque 2001}.
Most of the grains are in a ballistic motion between one vibrating
wall and the lid, or between two vibrated walls, following the selected
experimental configuration of the container. Thus, the low density
limit of a non-interacting granular matter is reached after a progressive
reduction of the dissipated internal energy. This reduction is due
to the decreasing frequency of the inelastic grain-grain collisions.

\begin{figure}
\includegraphics[%
  width=80mm,
  keepaspectratio]{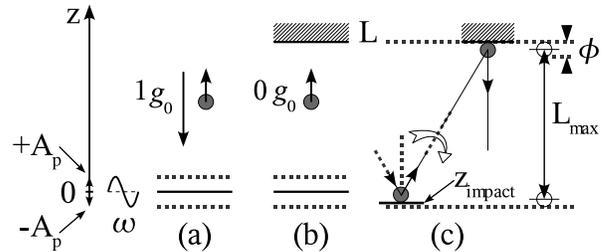}

\vspace{-3mm}
\caption{(a) Pustylnikov version of the Fermi acceleration, in which a ball
returns to an oscillatory wall under the $g_{0}$ Earth's gravity
acceleration (see \cite{Lichtenberg 1980,Lichtenberg 1992} and references
therein); (b) Ulam version, in which a ball bounces back and forth
between an oscillating wall and a fixed wall separated by distance
L under $0\, g_{0}$ (see \cite{Lichtenberg 1980,Holmes 1982} and
references therein); (c) schematic illustration of the ball rotation
and transverse velocity fluctuations frozen in the dissipative Ulam
version with restitution coefficient $\varepsilon<1$. The particle
motion appears then quasi-1D and regular, with a significant increase
of its average energy due to a non-stochastic acceleration when the
wall oscillation is a periodic function of time.}\vspace{-5mm}

\end{figure}

Our basic understanding of single particle dynamics comes from $1\, g_{0}$
experiments of one bouncing ball (diameter $\phi$) on a vibrating
plate, when the restitution coefficient $\varepsilon$ associated
with the ball-plate contact exhibits a finite value $0<\varepsilon\leq1$
\cite{Mehta 1990,Warr 1996,Geminard 2003}. The analysis of the results
is mainly concentrated on the rich phase space for the long-term behavior
of this impacting system, which makes questionable the experimental
conditions to observe the evolution to chaos \cite{Mehta 1990}. Thus,
our initial intuition suggests that the ball dynamics in a finite-sized
box remains poorly affected by a second wall, provided that the wall-plate
distance $L$ is larger than the vibration amplitude $A_{p}$. In
fact, as demonstrated in $1\, g_{0}$ and random-$low\left(<1\right)\, g_{0}$
experiments below, the assumption where stochastic trajectories of
the ball occur as the most probable dissipative situations for the
long-term behavior of the system, is not realistic \emph{within a
large plate velocity range}. On the contrary, the ball tends to behave
quasi-instantaneously as a regular particle in a 1D vibrating cavity
with the translation motion parallel to the vibration direction. In
the same time, a significantly audible sound is generated by the characteristic
resonant {}``impact noise''. The ball resonant behavior demonstrates
a drastic reduction of the phase space dimension because the second
wall increases the dissipation, eliminating the rotation of the ball
and freezing its transverse velocity-fluctuations after a very few
number of back and forth, as schematically shown on Fig. 1c and as
discussed below. The gain on the mean velocity of the resonant ball
is used here to report very accurate variations of the normal restitution
coefficient as a function of the ball velocity.

\begin{figure}[h]
\includegraphics[%
  width=80mm,
  keepaspectratio]{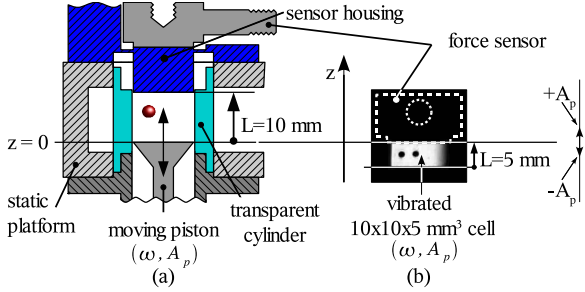}

\vspace{-3mm}
\caption{(a) Experimental set-up of the Ulam model for 1-ball ($\phi=2.0\, mm$
diameter), used for ground-based studies ($z$-axis parallel to Earth's
gravity direction) and for low-gravity studies (residual random acceleration
$\leq5\,10^{-2}\, g_{0}$) during parabolic flights of the CNES-A300-ZeroG
airplane. (b) Part of video picture showing 2-ball coherent flight
positions ($\phi=1.2\, mm$ ball diameter), vibrated along the $z$-axis
at $f=\frac{\omega}{2\pi}=14.75\, Hz$ and $A_{p}=0.3241\, mm$, in
very low-gravity conditions (residual random acceleration $\leq10^{-4}\, g_{0}$),
during our experiment on the ESA-funded sounding rocket Maxus 5 (see
text). The sensor location appears in dashed line.}\vspace{-3mm}

\end{figure}

Our first experiment studies the dynamics of a single ball in a static
cylindrical cell of $L=10\, mm$ height, closed at the bottom by a
vibrating piston (Fig. 2a). The piston moves along the $z$-axis,
according to $z=A_{p}sin\left[\omega t\right]$ (i.e. acceleration
$g_{z}=-\Gamma_{p}sin\left[\omega t\right]$ with $\Gamma_{p}=\omega^{2}A_{p}$).
$g_{z}$-accelerations are monitored using a piezoelectric tri-axial
accelerometer (PCB Piezotronics, Model M356A08) attached to the moving
part of an electromagnetic shaker. The piston, $D_{p}=12.7\, mm$
in diameter, is made from type AISI 316L stainless steel. The static
transparent cylinder, $13.0\, mm$ inner diameter, $20\, mm$ outer
diameter, and $22\, mm$ height, is made from PMMA. At the top of
the cell, the ball impacts the flat cylindrical cap ($12.7$ $mm$
diameter, $9.0$ $mm$ thickness) of a covered-sensor housing made
from type 17-4 stainless steel, in contact with the flat sensing surface
of a force sensor (PCB Piezotronics, Model 200B02). The ball resonant
motion is observed by stroboscopic illumination at the shaker frequency.
The signals from the $z$-axis accelerometer and the force sensor
are recorded with a resolution of $0.5\,\mu s$. Amplitude, frequency,
and acceleration ranges used here are $0.44\, mm<A_{p}<0.62\, mm$,
$30\, Hz\lesssim f=\frac{\omega}{2\pi}\lesssim120\, Hz$, and $3\, g_{0}<\left|g_{z}\right|<40\, g_{0}$,
respectively.

This vibrating facility was operated both on ground and under reduced-gravity
conditions, during parabolic flights on the French Space Agency (CNES)
A300 ZeroG airplane. One experimental run time covers a typical $\Delta t$
duration of $\simeq20$ $s$, (the low-gravity period of a parabolic
flight), during which the numerical data are stored.

\begin{figure}[h]
\includegraphics[%
  clip,
  width=80mm,
  keepaspectratio]{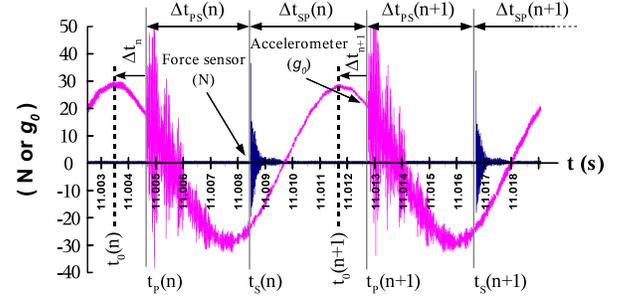}

\vspace{-3mm}
\caption{$1\, g_{0}$ synchronized signals of the $z$-axis accelerometer
and force sensor showing impact times (lower) and flight times (upper)
for back and forth resonant behavior of a single ball at $f=121.2\, Hz$
with $A_{p}=0.486\, mm$. For each period, $t_{0}\left(n\right)$
corresponds to the piston position $z\left[t_{0}\left(n\right)\right]=-A_{p}$,
where the length cavity is maximum (see text). }\vspace{-3mm}

\end{figure}

Fig. 3 reports typical signals for a ball resonant behavior in the
case of one stainless steel spherical ball, $\phi=2.000\pm0.002\, mm$
in diameter, vibrated at $f=121.2\, Hz$ with $A_{p}=0.486\, mm$
($\Gamma_{p}=28.7\, g_{0}$). The $z$-axis accelerometer response
to each ball-piston impact is superimposed on the sinusoidal variation
of $g_{z}$-component. This figure gives a typical sequence of $n$,
$n+1$ \emph{}resonant impacts on the piston (subscript $P$) and
on the force sensor cap (subscript $S$). The reference time $t_{0}(n)=\frac{3n}{4f}$
of the $n^{th}$ sequence starts at the time when the piston reaches
its maximum amplitude position for which the cavity length is maximum.
$\Delta t_{n}=t_{0}\left(n\right)-t_{P}\left(n\right)$ is the time
delay for the $n^{th}$ ball-piston impact at $t_{P}\left(n\right),$
where the piston position is $z_{impact}\left(n\right)=-A_{p}cos\left[\varphi_{n}\right]$,
associated to the impact phase $\varphi_{n}=-2\pi f\Delta t_{n}$.
The impact times, $t_{P}\left(n\right),$ $t_{S}\left(n\right)$,
the flight times, $\Delta t_{PS}\left(n\right)$, $\Delta t_{SP}\left(n\right)$,
the time delay $\Delta t_{n}$, the impact position $z_{impact}\left(n\right)$
and phase $\varphi_{n}$ of the piston, are then obtained by numerical
data analysis over each run time $\Delta t$. Our measurement precision
($\leq5\,\mu s$) of relative times appears to be much bigger than
in previous bouncing ball experiments \cite{Stensgaard 2001,Labous 1997}.

The ideal resonant behavior over $\Delta t$, gives $n_{T,ideal}=f\Delta t$
for the ideal total impact number, while our statistical signal analysis
counts the effective impact number $n_{T,eff}$. Fig. 4a gives the
ball resonance rate (\%), $\left(\frac{n_{T,eff}}{n_{T,ideal}}\times100\right)$,
as a function of $f$, for a nearly constant amplitude value $A_{p}\simeq0.5\, mm\simeq\frac{L}{20}\simeq\frac{\phi}{4}$.
The results from $1\, g_{0}$ (full diamonds) and $\leq5\,10^{-2}\, g_{0}$
(open diamonds) experiments are reported. Fig. 4a shows that the ball
resonant behavior is all the more frequent as $f$ increases and gravity
level decreases. This result enhances the relative influence of the
gravity effects and/or of the finite size effects at low frequency.

In addition, the $\varphi_{n}$ statistical analysis provides the
$\frac{\left\langle \left|z_{impact}\right|\right\rangle }{A_{p}}=cos\left[\left\langle \varphi\right\rangle \right]$
behavior as a function of $f$ which is reported on Fig. 4b ($\left\langle x\right\rangle $
corresponds to $x$ mean value). The resonance corresponds to the
nearly-maximal length of the cavity for which the gain on the ball
energy is then a (maximal) extremum, while the wall velocity at the
impact tends to a (minimal) extremum (close to zero). These results
can be understood in terms of the small but finite impact dissipation
of energy ($\sim1-\varepsilon^{2}$), with $\varepsilon<1$. The periodic
condition at $0\, g_{0}$ gives $v_{b,up}=\frac{1+\varepsilon_{P}}{1-\varepsilon_{P}\varepsilon_{S}}v_{p}$
and $v_{b,down}=-\frac{\left(1+\varepsilon_{P}\right)\varepsilon_{S}}{1-\varepsilon_{P}\varepsilon_{S}}v_{p}$,
where $v_{p}$, $v_{b,up}$, and $v_{b,down}$, are the respective
velocities for the piston, for the ball moving up from the piston
to the sensor cap, and down from the sensor cap to the piston. $\varepsilon_{P}$
and $\varepsilon_{S}$ are the respective restitution coefficients
for the ball-piston and ball-sensor cap contacts. The regular behavior
at fixed values of $L-\phi$, $\varepsilon_{P}$, $\varepsilon_{S}$,
$f$ and $A_{p}$, satisfies\begin{equation}
\alpha+cos\left[\left\langle \varphi\right\rangle \right]=\beta sin\left[\left\langle \varphi\right\rangle \right]\label{eq:1}\end{equation}
with $\alpha=\frac{L-\phi}{A_{p}}$ and $\beta=2\pi\frac{1+\varepsilon_{P}}{1+\varepsilon_{S}}\frac{\varepsilon_{S}}{1-\varepsilon_{P}\varepsilon_{S}}$.
This non-equilibrium steady state expresses the balance between forced
and dissipated energies, which occurs for one single mean reduced
length of the cell (given by the left hand member of Eq. (1)). That
corresponds to a single mean momentum exchange in the simplified Ulam
version \cite{Lichtenberg 1980,Holmes 1982,Lichtenberg 1992} for
$\varepsilon<1$, in which the moving wall imparts momentum to the
ball but occupies a fixed position.

\begin{figure}[h]
\includegraphics[%
  width=80mm,
  keepaspectratio]{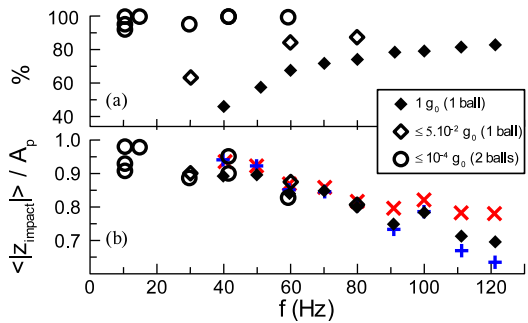}

\vspace{-3mm}
\caption{(a) Ball resonance rate (\%); (b) Relative wall position at the ball
impact time. Full diamonds) 1-ball $1\, g_{0}$ experiment; Open diamonds)
1-ball airplane experiment; Open circles) 2-ball Maxus-5 experiment.
In (b), black $+$ and grey $\times$ correspond to the $cos\left[\left\langle \varphi\right\rangle \right]$,
with the $\left\langle \varphi\right\rangle $-determination from
Eq. (1), for $\left\langle \varepsilon_{S}\right\rangle =\left\langle \varepsilon_{P}\right\rangle =\varepsilon$
mean values shown in Fig. 5, at each selected $\left\{ f;A_{p}\right\} $
run (note the agreement with direct analysis of $\frac{\left\langle \left|z_{impact}\right|\right\rangle }{A_{p}}$).
$\left|z_{impact}\right|$ corresponds to the position of the flat
top of the piston (1-ball experiment), or the flat cap of the force
sensor (2-ball Maxus-5 experiment).}\vspace{-3mm}

\end{figure}

Therefore, our accurate velocity measurements associated with the
regular behavior of the ball provide direct access to $\varepsilon_{S}=-\frac{v_{b,down}}{v_{b,up}}$
and $\varepsilon_{P}=\frac{v_{b,up}-v_{p}}{v_{p}-v_{b,down}}$. Fig.
5 shows the statistical behavior of $\varepsilon_{S}$ (black $+$),
and $\varepsilon_{P}$ (grey $\times$), as a function of $v_{i}=v_{b,down}$,
and $v_{i}=v_{b,up}$, respectively (accounting for gravity effects).
We observe that $\varepsilon_{P}\cong\varepsilon_{S}$, leading to
simplified forms of above relations. That permits to check the validity
of the Eq.(1), as shown by $cos\left[\left\langle \varphi\right\rangle \right]$
values (black $+$ and grey $\times$), reported in Fig. 4b, which
have been obtained using $\left\langle \varepsilon_{S}\right\rangle =\left\langle \varepsilon_{P}\right\rangle =\varepsilon$
mean values, (black $+$ and grey $\times$, within white square),
shown in Fig. 5, for each selected $\left\{ f,A_{p}\right\} $ pair.
We also note that our $\varepsilon$ measurements are in sharp contrast
with the solid curve representing a recent fit of former results \cite{McNamara 1999}
(see also Ref.\cite{Goldsmith 1960}). It highlights that only the
absence of significative ball rotation can explain such $\varepsilon$
high values.

\begin{figure}[h]
\includegraphics[%
  bb=0bp 0bp 90mm 45mm,
  width=80mm,
  height=40mm]{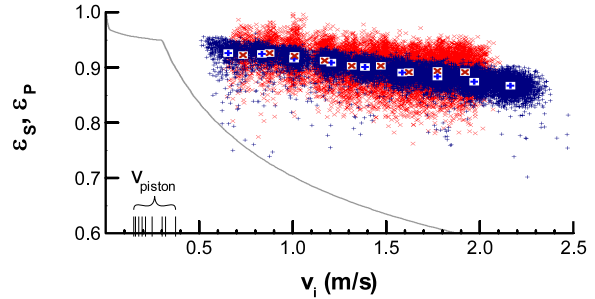}

\caption{$\varepsilon_{S}$ (black $+$), and $\varepsilon_{P}$ (grey $\times$),
measurements for ball-sensor cap and ball-piston contacts (see text),
according to ball incident velocity ($1\, g_{0}$ experiments). Symbols
in white squares correspond to associated mean values which are used
for$\left\langle \varphi\right\rangle $-determination from Eq. (1),
for each selected $\left\{ f;A_{p}\right\} $ run (see Fig. 4b). The
maximum velocity range for our vibrating piston is indicated on the
$v_{i}$-axis below the bracket. The full curve corresponds to a fit
\cite{McNamara 1999} of previous measurements. The decrease in restitution
at large velocity is interpreted as being due to plasticity effects
in the solid \cite{Goldsmith 1960}. }
\end{figure}

A second experiment was performed on the Maxus-5 sounding rocket funded
by the European Space Agency (ESA), where residual random acceleration
was $\leq10^{-4}\, g_{0}$. For the first time, this experiment studies
the dynamics of two \emph{}hard-brass spherical balls, $\phi=1.190\pm0.002\, mm$
in diameter, in a vibrating parallelepipedic box of $5\, mm$ height
and $10\times10\, mm^{2}$ internal cross section, where the sensitive
cap of the force sensor is used as an opposite cell-wall (see Fig.
2b for details). The experimental conditions maintain the ratio relation
$\frac{A_{p}}{L}\simeq\frac{1}{4}\frac{\phi}{L}$. Fig. 6a is similar
to Fig. 3, except that the sensor position and the ball-sensor impact
force are the only recorded signals. Fig. 6a evidences nearly ideal
concomitancy of the 2-ball resonant behaviors during a time period
of $500\, ms$ selected among the $65\, s$ run at $f=14.75\, Hz$,
with $A_{p}=0.3241\, mm$ ($\Gamma_{p}=0.3\, g_{0}$). The $\Delta t_{n}=t_{0}\left(n\right)-\frac{1}{2}\left[t_{S,1}\left(n\right)+t_{S,2}\left(n\right)\right]$
values for the $n^{th}$ 2-ball impacts occurring at $t_{S,1}$ and
$t_{S,2}$, respectively, are reported on Fig. 6b, for a total number
of recorded impacts ($2n_{T,eff}=1906$) very close to ideal value
($2n_{T,ideal}=1918$). This figure illustrates the quasi-perfect
coherent behaviors of the two balls during this long run time. This
coherent behavior is called hereafter the {}``laser-like'' behavior
for granular matter at very low density. Fig. 4a (open circles) shows
that this {}``laser-like'' behavior approaches the 100\% ball resonance
rate in weightlessness for a significative range of low frequencies,
confirming then the gravity-sensitivity of the ball acceleration mechanism.
Fig. 4b (open circles) extends the behavior of the (sensor) wall mean
position at impact times to the low frequency range.

\begin{figure}[h]
\includegraphics[%
  bb=0bp 0bp 180mm 140mm,
  clip,
  width=80mm,
  keepaspectratio]{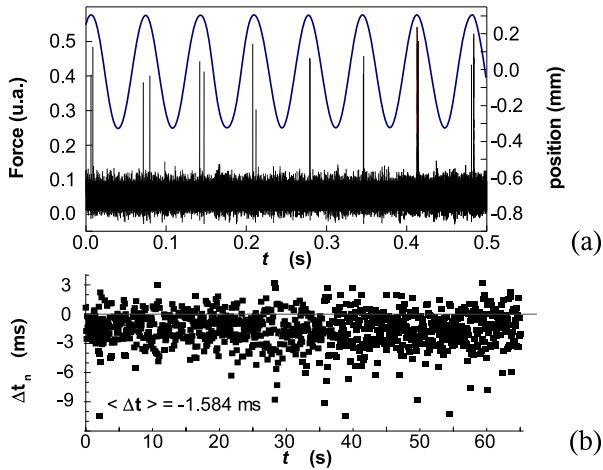}

\vspace{-2mm}
\caption{(a) low-gravity typical resonant behavior of two balls (ESA Sounding
rocket Maxus-5 experiment) for a period of $500\, ms$ selected among
a vibration run at $f=14.75\, Hz$ with $A_{p}=0.3241\, mm$ amplitude;
upper) $z$-position of the vibrated box ; lower) force sensor response
showing regular two ball impacts; (b) Variation of $\Delta t_{n,}$
(mod $f^{-1}$) along the $65\, s$ of the run, corresponding to $2n_{T,eff}=1906$
impacts of the 2 balls during their {}``laser-like'' behavior (see
text). Time reference is when position is maximum (see Fig. 2b and
text). The mean value is $\left\langle \bigtriangleup t\right\rangle =\left(n_{T,eff}\right)^{-1}\sum_{n}\Delta t_{n,}$.}\vspace{-2mm}

\end{figure}

We can approach the inelastic ball resonant behavior from a billiard-like
viewpoint where closed orbits are not ergodic, i.e. the so-called
eigen modes of the billiard cavity. The vibration excites the ball
motion on these modes. Those that dissipate too much cannot be sustained,
while those that dissipate too little do not exist or split precisely
on eigen modes. Therefore, in our present 3D-configuration of the
box, only a few eigen modes can be excited by vibrations. That explains
the few possible orbits (such as the observed one, parallel to the
vibration direction) which act as attractive basins with lowest dissipation.
Thanks to this phenomenon, the real shape of the cavity should play
a role in the ergodic/non-ergodic problem. However the ball orbit
parallel to the vibration direction remains stable, for example when
we tilt (up to 10°) the sensor cap surface compared to the perpendicular
direction of the vibration (which simulates a distorsion of the cavity
shape), or when we add a second ball of lower diameter (which simulates
an obstacle within the cavity). That confirms the loss of ergodicity
in this 3D experimental problem and the reduction of the non-interacting-ball
phase space by dissipation. Each 1-ball phase space typically goes
from a 11D space (3D for positions, 2D for rotations, 5D for associated
velocities, and 1D for time) to a 1D space (for time).

Considering the fact that this observed resonant behavior cannot correspond
to the case of the so-called Knudsen-like regime ($l_{g}\gg L$),
in which particles explore space ergodically but do not increase their
energy on the average, we conclude that our understanding of the very
low density limit of a non-interacting granular matter should be revisited,
in the absence of gravity, in order to investigate : i) the {}``low-energy''
regime, corresponding to small aspect ratio $\frac{v_{piston}}{v_{ball}}\approx\frac{A_{p}}{L-\phi}\ll1$,
which can permit one to check the relevance of a threshold value $\left(\frac{v_{piston}}{v_{ball}}\right)_{th}\lesssim1-\varepsilon$
needed to observe some more irregular motion; ii) the {}``high-energy''
regime, corresponding to larger values of the surface ratio $N\left(\frac{\phi}{D_{p}}\right)^{2}$,
for which we can expect that above a possible threshold number $N_{min}$
of balls (which remains to be determined), the presently observed
{}``laser-like'' behavior could be replaced by the classical dynamical
behavior where more frequent inelastic interparticle collisions increase
dissipation and restore the ergodic motion of a dilute granular {}``gas''.

We thank S. Fauve and E. Falcon for helpful discussions. This work
was supported by CNES and ESA. The authors gratefully acknowledge
Novespace and {}``Centre d'Essais en Vol'' teams for their assistance
during A300 Zero-G airplane experiments. The TEM-FER Maxus 5 experimental
module has been constructed by EADS Space Transportation (Germany).
We gratefully acknowledge the Maxus team for its technical assistance.

\end{document}